\documentclass[11pt,conference]{IEEEtran}
\usepackage{amsmath,amsfonts,cite}

\def\bdA{\boldsymbol{A}}
\def\bdB{\boldsymbol{B}}
\def\bdH{\boldsymbol{H}}
\def\bdI{\boldsymbol{I}}
\def\bdP{\boldsymbol{P}}
\def\bdQ{\boldsymbol{Q}}
\def\bdU{\boldsymbol{U}}
\def\bdV{\boldsymbol{V}}
\def\bdX{\boldsymbol{X}}
\def\bdY{\boldsymbol{Y}}
\def\bdZ{\boldsymbol{Z}}
\def\bdx{\boldsymbol{x}}
\def\bdzeros{\boldsymbol{0}}
\def\bdones{\boldsymbol{1}}
\def\E{\text{E}}
\def\mD{\mathcal {D}}
\def\bydef{:=}
\def\M#1{M_{#1}}
\def\N#1{N_{#1}}
\def\CN#1{\mathcal{CN}(#1)}
\def\R#1{R_{#1}}

\def\rank{\text{rank}}

\newtheorem{theorem}{Theorem}
\newtheorem{lemma}{Lemma}
\def\secref#1{Section~\ref{#1}}
\def\lemref#1{Lemma~\ref{#1}}

\makeatletter\def\@oddfoot{\hfill\thepage\hfill}\makeatother
\normalsize

\begin{document}

\title{On the Degrees of Freedom Regions of Two-User MIMO Z and Full
Interference Channels with Reconfigurable Antennas }

\author{
\IEEEauthorblockN{\normalsize Lei Ke and Zhengdao Wang}
\IEEEauthorblockA{Dept. of ECE, Iowa State University, Ames, IA
50011, USA. \\Email: \{kelei, zhengdao\}@iastate.edu}}

\maketitle

\begin{abstract}
We study the degrees of freedom (DoF) regions of two-user multiple-input
multiple-output (MIMO) Z and full interference channels in this paper. We
assume that the receivers always have perfect channel state information. We
derive the DoF region of Z interference channel with channel state information
at transmitter (CSIT). For full interference channel without CSIT, the DoF region
has been obtained in previous work except for a special case $\M1<
\N1<\min(\M2,\N2)$, where $M_i$ and $N_i$ are the number of transmit and
receive antennas of user $i$, respectively. We show that for this case the DoF
regions of the Z and full interference channels are the same. We establish the
achievability based on the assumption of transmitter antenna mode switching. A
systematic way of constructing the DoF-achieving nulling and beamforming
matrices is presented in this paper.

\end{abstract}

\section{Introduction} Characterizing capacity region of interference channel
has been a long open problem. Many researchers investigated this important
area, and the capacity regions of certain interference channels are known when
the interference is strong, e.g. \cite{carl75,sato81,gaco82}. However, when
the interference is not strong, the capacity region is still unknown. Recent
progress reveals the capacity region for two-user interference channel within
one bit \cite{ettw08}, and after that the capacity region for very weak
interference channel is settled \cite{shkc08c,srve08c,mokh08c}. Recently, a
deterministic channel model has been proposed and used to explore the capacity
of Gaussian interference network \cite{gbdt08,gbpt08,sadt09} such that the gap
to capacity region can be bounded up to a constant value.

When it comes to multiple-input multiple-output (MIMO) networks,
the capacity regions of certain MIMO interference channels are
known \cite{avvv09,sckp09}. Instead of trying to characterize the capacity
region completely, the degrees of freedom (DoF) region characterizes how
capacity scales with transmit power as the signal-to-noise ratio goes to
infinity.

It is well-known that in certain cases, the absence of channel state
information at transmitter (CSIT) will not affect the DoF
for MIMO networks, e.g., in the multiple access channel \cite{gjjv03}. In
other cases, CSIT does play an important role. For example, using interference
alignment scheme, it is shown that the total DoF of a $K$-user MIMO
interference channel can be $MK/2$, where $M$ is the number of antennas of
each user \cite{caja08}. The key idea is to pack interferences from multiple
sources to reduce the dimensionality of signal space spanned by interference.

The DoF region of two-user MIMO interference channel with CSIT has been
obtained \cite{jafa07}, where it is shown that zero forcing is enough to
achieve the DoF region. However, it is a different story in two-user MIMO X
channel, where each transmitter has a message to every receiver. In
\cite{jash08} it is shown that interference alignment is the key to achieving
the DoF region of MIMO X network. The DoF region of two-user MIMO broadcast
channel and interference channel without CSIT are considered in \cite{hjsv09},
where there is an uneven trade-off between the two users.
Except for a special case, the DoF region for the interference channel is
known and achievable. Similar, but more general result of isotropic fading
channel can be found in \cite{yzdg09c}. The DoF regions of the $K$-user MIMO
broadcast, interference and cognitive radio channels are derived in
\cite{cvmv09} for some cases. However, the special case in \cite{hjsv09}
remains unsolved.

When only one of the two transmitter-receiver pairs is subject to
interference, then the interference channel is termed as Z interference
channel (ZIC). To avoid confusion, we will call the channel where both pairs
are subject to interference the full interference channel (FIC). The capacity
regions of MIMO Gaussian ZIC is established in \cite{sckp09c} under very
strong interference and aligned strong interference cases. In \cite{yzdg09},
the authors considered the capacity region of a single antenna ZIC without
CSIT using deterministic approach.

Recently, it is shown in \cite{jafar09} that if the channel is staggered block
fading, we can explore the channel correlation structure to do interference
alignment, where the upper bound in the converse can be achieved in some
special cases. For example, it is shown that for two-user MIMO  staggered block
fading FIC with 1 and 3 antennas at transmitters, 2 and 4 antennas at their
corresponding receivers and without CSIT, the DoF pair $(1, 1.5)$ can be
achieved. The idea was further clarified in \cite{cwtgs10}, where a blind
interference alignment scheme is also proposed for $K$-user multiple-input
single-output (MISO) broadcast channel to achieve DoF outer bound when CSIT is
absent. Their approach is to use reconfigurable antennas such that the
antennas can be switched among different modes to artificially create channel
variation. It is their work that inspires us to investigate the remaining
unknown case in the two-user MIMO FIC.

In this paper, we investigate the DoF region of MIMO ZIC with or without CSIT.
We found that for the unknown cases of MIMO FIC it is enough to consider an
equivalent ZIC, for which we propose a joint beamforming and nulling scheme to
achieve its converse upper bound. We assume that perfect CSI is always
available at receivers and transmitter of one user is capable of antenna mode
switching. Based on this, we describe the DoF regions of MIMO two-user ZIC and
FIC.

We first present the system model in \secref{sec.model}. The known results on
the DoF region of two-user MIMO FIC are briefly reviewed in
\secref{sec.ic}. We present the exact DoF regions of ZIC and FIC in
\secref{sec.zic}. Finally \secref{sec.conc} concludes our results.

Notation: boldface uppercase letters denote matrices or vectors.
$\mathbb{R},\mathbb{C}$ are the real and complex numbers sets.
$\CN{0,1}$ denotes a circularly symmetric complex normal
distribution with zero mean and unit variance. We use $\bdA\otimes
\bdB$ to denote the Kronecker product between $\bdA$ and $\bdB$.
$\bdzeros$ and $\bdones$ denote all one and all zero vectors,
respectively. $\bdA^T$ and $\bdA^H$ denote the transpose and
Hermitian of $\bdA$, respectively. We also use notation like
$\bdA_{m\times n}$ to emphasize that $\bdA$ is of size $m \times n$.
We use $\bdI_m$ to denote a size $m\times m$ identity matrix.

\section{System Model}\label{sec.model} Consider a MIMO interference channel
with two transmitters and two receivers, the number of transmit (receive)
antennas at the $i$th transmitter (receiver) is denoted as $M_i$ ($N_i$),
$i\in \{1,2\}$. The system is termed as an ($\M1,\N1,\M2,\N2$) system, which
can be described as
\begin{align}
\bdY_1(t)=\bdH_{11}(t)\bdX_1(t)+\bdH_{12}(t)\bdX_2(t)+\bdZ_1(t)\\
\bdY_2(t)=\bdH_{21}(t)\bdX_1(t)+\bdH_{22}(t)\bdX_2(t)+\bdZ_2(t)
\end{align}
where $t$ is the time index, $\bdY_i(t)\in \mathbb{C}^{N_i}$, $\bdZ_i(t)\in
\mathbb{C}^{N_i}$ are the received signal and additive noise of receiver $i$,
respectively. The entries of $\bdZ_i(t)$ are independent and identically
$\CN{0,1}$ distributed in both time and space. The channel between the $i$th
transmitter and the $j$th receiver is denoted as $\bdH_{ji}(t)\in
\mathbb{C}^{N_j \times M_i}$. We assume that the probability of $\bdH_{ij}(t)$
belonging to any subset of $\mathbb{C}^{N_j \times M_i}$ that has zero
Lebesgue measure is zero. For the two-user MIMO ZIC, $\bdH_{21}(t)=0$.
$\bdX_i(t) \in \mathbb{C}^{M_i}$ is the input signal at transmitter $i$ and
$\bdX_1(t)$ is independent of $\bdX_2(t) $. The transmitted signals satisfy
the following power constraint:
\begin{align}
\E(||\bdX_i(t)||^2)\leq P \quad i=1,2
\end{align}

Denote the capacity region of the two-user MIMO system as $C(P)$, which
contains all the rate pairs $(R_1,R_2)$ such that the corresponding
probability of error can approach zero as coding length increases. The DoF
region is defined as follows \cite{hjsv09}
\begin{multline*}
\mD \bydef \left\{(d_1,d_2)\in \mathbb R^+_2: \exists
(\R1(P),\R2(P))\in C(P),\right. \\
\left. \text{such that } d_i
=\lim_{P\to\infty}\frac{R_i(P)}{\log(P)}  , \quad i=1,2 \right \}.
\end{multline*}

\section{Known Results on DoF Region of MIMO Full Interference Channel}
\label{sec.ic}

We first present some known results on DoF region of MIMO full interference
channel which is useful for developing our results.

The degrees of freedom region of two-user MIMO full interference channel with
CSIT is the following \cite[Theorem 2]{jafa07}:
\begin{align}
d_i &\leq \min(M_i,N_i), \quad i=1,2; \\
d_1+d_2&\leq \min(\max(N_1,M_2),\max(M_1,N_2),\nonumber \\ &\qquad\qquad\N1+\N2,\M1+\M2)
\end{align}

An outer bound of degrees of freedom region of two-user MIMO full interference
channel without CSIT is as follows \cite[Theorem 1]{yzdg09c}:
\begin{align}
\text{for } i=1,2,\quad d_i &\leq \min(M_i,N_i);  \label{eq.nocsit.upper0}\\
d_1 \!+\!\frac{\min(\N1,\N2,\M2)}{\min(\N2,\M2)}d_2 & \leq \min(M_1+M_2,N_1);
        \label{eq.nocsit.upper}\\
\frac{\min(\N1,\N2,\M1)}{\min(\N1,\M1)}d_1\!+\!d_2 & \leq \min(M_1+M_2,N_2).
        \label{eq.nocsit.upper1}
\end{align}
Note that the same result is also given in \cite{hjsv09}, though in a less
compact form.

It is known \cite{{hjsv09}} that when $\N1<\N2$, the outer bound given in
\eqref{eq.nocsit.upper0}, \eqref{eq.nocsit.upper} and \eqref{eq.nocsit.upper1}
can be achieved by zero forcing or time sharing except for the case $\M1 < \N1
< \M2$, for which we do not know how to achieve
\begin{align}(d_1,
d_2)=(\M1,\frac{\min(\M2,\N2)(\N1-\M1)}{\N1}) \label{eq.DoFunknown}
\end{align}
in general.

The outer bound given in \cite[Theorem 1]{yzdg09c} is derived based on the
assumption that i) $\bdH_{12}$ and $\bdH_{22}$ are statistically equivalent,
and ii) $\bdH_{21}$ and $\bdH_{11}$ are statistically equivalent. Since these
assumptions still hold when antenna mode switching is used at transmitters,
the DoF outer bound specified by
\eqref{eq.nocsit.upper0}--\eqref{eq.nocsit.upper1} is still valid.

\section{DoF Regions of MIMO ZIC and FIC }\label{sec.zic}

In this section we will first discuss the DoF regions of MIMO ZIC with or
without CSIT, the extension to MIMO FIC will be made at the end of this
section.

\subsection{Two-User MIMO ZIC with CSIT}

\begin{theorem}
\label{thm.z.csit.DoF}The degrees of freedom region of two-user MIMO Z
interference channel with CSIT is the following
\begin{align*}
d_i &\leq \min(M_i,N_i), \quad i=1,2, \\
d_1+d_2&\leq \min(\max(N_1,M_2),\N1+\N2,\M1+\M2).
\end{align*}
\end{theorem}
\begin{IEEEproof} The theorem can be  proven based on the
result in \cite{jafa07}. We omit the proof due to space limit.
\end{IEEEproof}

\subsection{Two-User MIMO ZIC without CSIT}

\begin{lemma}
\label{thm.nocsit.zDoF} The outer bound of degrees of freedom region of
two-user MIMO Z interference channel without CSIT can be given as
\begin{align}
&d_i \leq \min(M_i,N_i), \quad i=1,2 \label{eq.outer.z0} \\
&d_1 \!+\!\frac{\min(\N1,\N2,\M2)}{\min(\N2,\M2)}d_2\!\leq\!
        \min(M_1\!+M_2,N_1).
        \label{eq.outer.z}
\end{align}
\end{lemma}

\begin{IEEEproof}
This is the direct result of \cite[Theorem 1]{yzdg09c} by noticing that there
is no interference from transmitter 1 to receiver 2 hence
\eqref{eq.nocsit.upper1} is not longer needed.
\end{IEEEproof}

\begin{lemma}
\label{lem.znocsit.n1ln2}For the two-user MIMO Z interference channel,
when $N_1 \geq\N2$, \eqref{eq.outer.z} is achievable by zero forcing.
\end{lemma}
\begin{IEEEproof} This can be shown by noticing that  \eqref{eq.outer.z}
is reduced to
\(
d_1+d_2\leq\min(M_1+M_2,N_1)
\)
and zero forcing suffices.
\end{IEEEproof}

We have the following lemma regarding the relationship between DoF regions of ZIC
and FIC.

\begin{lemma}
\label{lem.ifcziceq} When $\N1\le \N2$, the MIMO ZIC and FIC have the
same DoF regions. Any encoding scheme that is DoF optimal for one channel is
also DoF optimal for the other.
\end{lemma}
\begin{IEEEproof} Any point in the FIC is also trivially achievable in the ZIC
because user 2's channel is interference free. Conversely, any point
achievable in the ZIC region, is also achievable in FIC. This is based on the
fact that the channels are statistically equivalent at both receivers. If
receiver 1 can decode user 1's message, then receiver 2, having at least as
many antennas, must also be able to decode the same message. Receiver 2 can
then subtract the decoded message, which renders the resulting channel the
same as in the ZIC.
\end{IEEEproof}

Due to \lemref{lem.ifcziceq}, we can translate all achievability schemes from
FIC to ZIC in a trivial way. However, for the case $\M1 < \N1 <
\min(\M2,\N2)$, there is still no DoF-optimal encoding scheme for the two-user
MIMO FIC.

In the following, we develop a scheme to achieve \eqref{eq.DoFunknown} for
MIMO ZIC, which is our main result of this paper.

Consider the two-user MIMO ZIC with $\M1 < \N1 < \M2=\N2$. We want to show
that the following DoF pair is achievable
\begin{align}(d_1,
d_2)=\left(\M1,\frac{\M2(\N1-\M1)}{\N1}\right).
\end{align}

We notice that this point can not be achieved by zero forcing over one time
instant. This is because using zero forcing if transmitter 1 sends $\M1$
streams, transmitter 2 can only send $\N1-\M1$ streams without interfering
receiver 1. If transmitter 2 sends more streams, the desired signal and
interference are not separable at receiver 1 as transmitter 2 does not know
channel state information so it can not send streams along the null space of
$\bdH_{12}$. A simple example is the $(1,2,3,3)$ case, where the outer bound
gives us $(1,1.5)$, which is not achievable via zero forcing over one time
slot.

We make the assumption that the channel $\bdH_{12}$ stays the same for at
least $\N1$ time slots, and there are $\N1$ antenna modes transmitter 1 can
use. It is sufficient to show that $(\M1\N1,\M2(\N1-\M1)) $ streams can be
achieved in $\N1$ time slots. The transmitter 1 will use different antenna
modes to create channel variation. The time expansion channel between
transmitter 1 and receiver 1 will be
\begin{align}
\tilde \bdH_{11}\!\!=\!\!\left[\! \begin{array}{ccccc}
\!\bdH_{11}(1) & \bdzeros & \bdzeros  & \bdzeros \\
\bdzeros & \!\bdH_{11}(2) & \bdzeros  & \bdzeros \\
\vdots & \vdots & \ddots & \vdots \\
\bdzeros & \bdzeros & \bdzeros  &\!\bdH_{11}(\N1)\! \\
\end{array} \right]_{\!\N1^2\! \times\! \N1\M1}\nonumber
\end{align}
and the channel between transmitter 2 and receiver 1 is
\begin{align}
\tilde \bdH_{12}=\bdI_{\N1}\otimes \bdH_{12}(1)
\end{align}
as user 2 does not create channel variation. Here and after, we use tilde
notation to indicate the time expansion signals. We will use precoding at
transmitter 2 only and nulling at receiver 1 only. Let $\tilde \bdP$ be the
transmit beamforming matrix at transmitter 2 and $\tilde \bdQ$ be the nulling
matrix at receiver 1. We propose to use the following structures for them
\begin{align}
\tilde \bdP_{\M2\N1 \times \M2(\N1-\M1)}&=\bdP_{\N1 \times
        (\N1-\M1)} \otimes \bdI_{\M2}\\
\tilde \bdQ_{\M1\N1 \times \N1^2}&=\bdQ_{\M1 \times \N1} \otimes \bdI_{\N1}.
\end{align}

The received signal at receiver 1 can be written as
\begin{align}
\tilde \bdY_{1}=\tilde \bdH_{11} \tilde \bdX_1+\tilde \bdH_{12} \tilde \bdP \tilde \bdX_2
+\tilde \bdZ_1,
\end{align}
where $ \tilde \bdX_1$ is a length $\M1\N1$ vector, and $ \tilde \bdX_2$ is a
length $\M2(\N1-\M1)$ vector. After applying nulling matrix $\tilde \bdQ$ we
have
\begin{align}
\tilde \bdQ\tilde \bdY_{1}=\underbrace{\tilde \bdQ\tilde \bdH_{11}}_{\tilde\bdU} \tilde \bdX_1+\underbrace{\tilde \bdQ\tilde \bdH_{12} \tilde \bdP}_{\tilde\bdV}  \tilde \bdX_2
+\tilde \bdQ\tilde \bdZ_1.
\end{align}
To achieve the degrees of freedom $(\M1\N1,\M2(\N1-\M1))$ for both users, it
is sufficient to design our $\tilde \bdP$ and $\tilde \bdQ$ to satisfy the
following conditions simultaneously
\begin{enumerate}
\item $\rank(\tilde\bdU)=\M1\N1$.
\item $\rank(\tilde \bdP)=\M2(\N1-\M1)$.
\item $\tilde \bdV=\bdzeros$.
\end{enumerate}
The second condition can be easily satisfied. Because $\rank(\tilde
\bdP)=\rank(\bdP) \rank(\bdI_{\M2})$, we only need to design $\bdP$ such that
$\rank(\bdP)=\N1-\M1$. As to the third condition, notice that
\begin{align}
\tilde \bdV&=(\bdQ\otimes \bdI_{\N1})(\bdI_{\N1}\otimes \bdH_{12}(1))(\bdP \otimes \bdI_{\M2})\\
&=(\bdQ\bdI_{\N1}\bdP)\otimes (\bdI_{\N1}\bdH_{12}(1) \bdI_{\M2})\\
&=\bdQ\bdP\otimes \bdH_{12}(1).
\end{align}
It is therefore sufficient (and also necessary) to have $\bdQ\bdP=0$. Then the
key problem is to find a $\bdQ$ such that
\begin{align}
\tilde\bdU=(\bdQ \otimes \bdI_{\N1} )\tilde\bdH_{11}
\end{align}
has full rank $\M1\N1$. The matrix $\tilde \bdU$ is of size ${\M1\N1 \times
\M1\N1}$ and has the following structure
\begin{align}
\!\!\!\tilde \bdU \!\! =\!\!\left[ \begin{array}{ccccc}
\!\!\!q_{11}\bdH_{11}(1)\!\!\! & \!\!\!q_{12}  \bdH_{11}(2)\!\!\! &\!\!\! \cdots  & \!\!\!q_{1N_1}  \bdH_{11}(\N1)\!\!\!\\
\!\!\!q_{21}\bdH_{11}(1)\!\!\! & \!\!\!q_{12}  \bdH_{11}(2)\!\!\! &\!\!\! \cdots  & \!\!\!q_{2N_1}  \bdH_{11}(\N1)\!\!\!\\
\vdots & \vdots & \!\!\!\ddots & \vdots \\
\!\!\!q_{\M11}\bdH_{11}(1) & \!\!\!q_{\M12}  \bdH_{11}(2) & \!\!\!\cdots  & \!\!\!q_{M_{1}N_1}  \bdH_{11}(\N1)\!\!\!
\end{array} \right]. \nonumber
\end{align}

To show that $\tilde \bdU$ has full rank, we use an idea of
\cite{sibu02} and need the following lemma.

\begin{lemma}
\cite[Lemma 1]{sibu02}\label{lem.fullrank} Consider an
analytic function $f(\bdx)$ of several variables $\bdx=[x_1,\dots,x_n]^T$. If
$f$ is nontrivial in the sense that there exists $\bdx_0\in\mathbb{C}^n$ such
that $f(\bdx_0)\neq 0$, then the zero set of $f(\bdx)$
\(
\mathrm{Z}:=\{\bdx\in \mathbb{C}^n | f(\bdx)=0 \}
\)
is of measure (Lebesgue measure in $\mathbb{C}^n$) zero. \hfill \QED
\end{lemma}

Because the determinant of $\tilde \bdU$ is an analytic polynomial function of
elements of $\bdH_{11}(t), t=1,\dots,\N1$, we only need to find a specific
pair of $\bdQ$ and $\bdH_{11}(t), t=1,\dots,\N1$ such that $\tilde \bdU$ is
full rank. We propose the following:
\begin{multline}
q_{mn}=\exp\left(-j\frac{2\pi(m-1)(n-1)}{N_1}\right) \\
        m=1,\cdots, M_1 \quad n=1,\dots,\N1. \label{eq.null}
\end{multline}
Let $W\bydef \exp(-j{2\pi}/{N_1^2})$. Take the realizations of $\tilde
\bdH_{11}(t)$, $t=1,\dots N_1$ as shown in \eqref{eq.hhh}.
\begin{figure*}
\begin{align}
 \bdH_{11}(t)=\left[ \begin{array}{ccccc}
W^0 & W^0 & \cdots  & W^0 \\
W^{t-1} & W^{N_{1}+t-1} & \cdots  & W^{(M_{1}-1)N_{1}+t-1} \\
\vdots & \vdots & \ddots & \vdots \\
W^{(N_1-1)(t-1)} & W^{(N_{1}-1)(N_{1}+t-1)} & \cdots  & W^{(N_{1}-1)[(M_{1}-1)N_{1}+t-1]} \\
\end{array} \right]_{\N1\times \M1} \label{eq.hhh}
\end{align}
\end{figure*}

It can be verified that for such choices of $\bdQ$ and $\bdH_{11}(t)$, $\tilde
\bdU$ is a Vandermonde matrix of different columns, and hence of full rank.
We also notice that $\tilde \bdU$ is a leading principal minor of a  permuted
fast Fourier transform (FFT) matrix with size $\N1^2\times \N1^2$. The
permutation is as follows: Index the columns of an FFT matrix $0,1,\dots,
N_1^2-1$, and then permute them in an order shown below:
\begin{multline*}
(0,N_1,2N_1,\dots,(M_1-1)N_1),\\ (1,N_1+1,2N_1+1, \dots, (M_1-1)N_1+1),...
\end{multline*}

Based on \lemref{lem.fullrank}, if we choose the nulling
matrix using $\bdQ$ as specified in \eqref{eq.null}, $\tilde\bdU$ has full
rank almost surely. One choice of the corresponding $\bdP$ matrix with respect
to \eqref{eq.null} is the following
\begin{multline}
p_{nm}=\exp\left(j\frac{2\pi(m-1)(n-1)}{N_1}\right), \\
        n=1,\dots,\N1, \quad m=\M1+1,\cdots, N_1 \label{eq.bftx2}.
\end{multline}
We note that the matrix $[\bdQ^H\; \bdP]$ is an FFT matrix. Our choices of
$\bdQ$, $\bdP$ in \eqref{eq.null} and \eqref{eq.bftx2} have a frequency domain
interpretation. The signal of user 2 is transmitted over frequencies
corresponding to the last $\N1-\M1$ columns of an FFT matrix, whereas the
first user's signal is transmitted on all frequencies. We also point out that
when $\N1/\M1=L\in\mathbb{Z}$, it is possible to achieve the same DoF using
only $L$ fold time expansion. Due to space limit, we will report that
elsewhere. The previous results can be summarized as:
\begin{lemma} \label{lem.dof.special}
For the two-user MIMO Z interference channel without CSIT, with antenna
numbers $\M1<\N1<\min(\M2,\N2)$. User 1 and user 2 can achieve DoF pair
$(\M1,\min(\M2,\N2)(\N1-\M1)/\M2)$. \hfill \QED
\end{lemma}

The achievability of \lemref{lem.dof.special} is based on time expansion,
antenna mode switching at transmitter 1, with jointly designed beamforming
matrix at transmitter 2 and nulling matrix at receiver 1. We end the
discussion of MIMO ZIC with the following theorem:

\begin{theorem}
The DoF region of two-user MIMO Z interference channel without CSIT is
described by the inequalities \eqref{eq.outer.z0} and \eqref{eq.outer.z} if
transmitter 1 has the antenna mode switching ability.
\end{theorem}
\begin{IEEEproof}
The achievability when $\N1 \geq \N2$ can be established by
\lemref{lem.znocsit.n1ln2}. When $\M1<\N1<\min(\M2,\N2)$, the achievability
follows from \lemref{lem.dof.special}. The achievability of the remaining
cases is based on the achievability of FIC given in \cite{hjsv09}.
\end{IEEEproof}

\subsection{Two-User MIMO FIC without CSIT} Finally, we have the following
result.

\begin{theorem}
The DoF region of two-user MIMO full interference channel without CSIT is
described by the inequalities \eqref{eq.nocsit.upper0},
\eqref{eq.nocsit.upper} and \eqref{eq.nocsit.upper1} if transmitter 1 has the
antenna mode switching ability.
\end{theorem}

\begin{IEEEproof} When $\M1<\N1<\min(\M2,\N2)$, we can achieve $(\M1,\min(\M2,\N2)(\N1-\M1)/{\M2})$
in the ZIC with same number of antennas. Hence, based on
\lemref{lem.ifcziceq}, it is also achievable in FIC. The achievability of the
remaining cases is given in \cite{hjsv09}.
\end{IEEEproof}

\section{Conclusions}\label{sec.conc}

We derived a few results on the exact DoF region for the MIMO Z and full
interference channels with perfect channel state information at receiver,
including results for i) the Z interference channel with and without channel
state information at the transmitter, and ii) the full interference channel
without channel state information at the transmitter. The achievability scheme
we designed for the case of $\M1 < \N1 < \min(\M2,\N2)$ relies on time
expansion, antenna mode switching at transmitter one, and has a frequency
domain interpretation. By combining our results with previously known results,
we completely characterized the DoF regions for both Z and full interference
channels when transmitter antenna mode switching is allowed. We comment that
when antenna mode switching is not allowed, the problem of DoF regions for
these channels are still not completely resolved.

\linespread{.94}\normalsize
\bibliographystyle{IEEEtran}
\bibliography{refs}

\end{document}